\documentclass[journal,draftcls,onecolumn,12pt,twoside]{IEEEtran}

\usepackage{graphicx} 
\usepackage{booktabs}
\usepackage{amsmath}
\usepackage{mathrsfs}
\usepackage{amsfonts}
\usepackage{amssymb}
\usepackage{epstopdf}
\usepackage{dblfloatfix}
\usepackage[caption=false,font=footnotesize]{subfig}
\usepackage{color}
\usepackage{enumitem}
\usepackage{amsthm}
\usepackage{algorithm,algorithmic}
\usepackage{multirow} 
\usepackage{array}
\usepackage{cite}
\usepackage{comment}
\usepackage{dsfont}
\usepackage{vmr-symbols-vecbold}

\def\cdf(#1)(#2)(#3){0.5*(1+(erf((#1-#2)/(#3*sqrt(2)))))}%

\providecommand{\ee}[1]{\exp\mathopen{}\left(#1\right)}

\providecommand{\e}[1]{e^{#1}}

\newcommand{\farg}[1]{\mathopen{}\left( #1 \right)}

\newcommand{\lefto}{\mathopen{}\left}
\newcommand{\vect}[1]{\boldsymbol{\mathbf{#1}}}

\begin{document}

\title{A Statistical Learning Approach to  Ultra-Reliable Low Latency Communication}

\author{Marko~Angjelichinoski, ~\IEEEmembership{Student~Member,~IEEE,}
      Kasper~Fl\o e~Trillingsgaard ~\IEEEmembership{Student~Member,~IEEE,}
     and~Petar~Popovski, ~\IEEEmembership{Fellow,~IEEE}
\thanks{The authors are with the Department of Electronic Systems, Aalborg University, Denmark (e-mail: $\left\{\mbox{maa,kft,petarp}\right\}$@es.aau.dk)}}


\markboth{IEEE Transactions on Communications (2018, submitted)}%
{Angjelichinoski et al. The Statistics of URLLC}

\maketitle

\begin{abstract}
Mission-critical applications require Ultra-Reliable Low Latency  (URLLC) wireless connections, where the packet error rate (PER) goes down to $10^{-9}$. Fulfillment of the bold reliability figures becomes meaningful only if it can be related to a statistical model in which the URLLC system operates. However, this model is generally not known and needs to be learned by sampling the wireless environment. In this paper we treat this fundamental problem in the simplest possible communication-theoretic setting: selecting a transmission rate over a dynamic wireless channel in order to guarantee high transmission reliability. We introduce a novel statistical framework for design and assessment of URLLC systems, consisting of three key components: (i) channel model selection; (ii) learning the model using training; (3) selecting the transmission rate to satisfy the required reliability. As it is insufficient to specify the URLLC requirements only through PER,   two types of statistical constraints are introduced, Averaged Reliability (AR) and Probably Correct Reliability (PCR).
The analysis and the evaluations show that adequate model selection and learning are indispensable for designing {consistent} physical layer that asymptotically behaves as if the channel was known perfectly, while maintaining the reliability requirements in URLLC systems.
\end{abstract}


\IEEEpeerreviewmaketitle

\section{Introduction}

\IEEEPARstart{U}{ltra}-reliable low latency communication (URLLC) is among the most exciting novelties in 5G networks \cite{Popovski2014,PopovskiCNTURLLC2018,Schulzetall2017}.
The projected reliability guarantees of $10^{-5}$, and even going down to $10^{-9}$, represent the ultra-reliable (UR) regime of wireless operation. This regime is necessary to support a multitude of mission-critical applications, such as remote control of robots, autonomous coordination among vehicles as well as many yet-to-be-defined use cases.
The strictness of the reliability requirements increases the focus on the performance of the \emph{physical layer}, its main design assumptions \cite{Popovski2014,Bennis2018}, the modeling of the inherently unreliable wireless channel \cite{Eggers2017,Swamy2018} and the adequacy of existing transmission strategies in UR-relevant regime \cite{Swamy2018}.

Fulfillment of the bold figures on reliability for URLLC becomes meaningful only if it can be related to a statistical model in which the URLLC system operates, as in that case one can calculate the probability of error or failure. However, the statistical model and the probability distribution of the parameters that affect the occurrence of errors are, generally, unknown and the URLLC system needs to invest resources to learn them. In general, the statistical model can encompass the interference from other devices, behavior of the protocols, etc. 
Here we consider the simplest possible communication-theoretic setting: selecting a transmission rate over a dynamic wireless channel, in absence of interference, in order to guarantee high transmission reliability. For example, even if it is known that a channel is subject to 
 a Rayleigh fading, guaranteeing certain reliability means that the average gain of that channel is known perfectly.

In this paper we address two fundamental questions: 
\begin{enumerate}
\item \emph{What level of knowledge is required about the wireless channel to be able to guarantee ultra-reliable support of a certain transmission rate?} To the best of our knowledge, the existing URLLC studies select the transmission parameters by ignoring the uncertainty introduced by the transmission environment. As a result, the assumption of perfect knowledge of the channel statistics needs to be revised, as without it, the extreme reliability guarantees become questionable. This assumption is fundamentally \emph{unreliable}, and although it might be acceptable in existing wireless systems with reliability targets in the order of $10^{-3}$, the impact of channel uncertainty will lead to severe performance degradation in URLLC systems.
\item \emph{How to pose the requirements for ultra-reliability in a statistically correct way?} Stating only that we need a packet error rate of e.g. $\epsilon=10^{-6}$ is insufficient. We define two different way to measure reliability: \emph{Averaged Reliability (AR)}, suitable for dynamically changing environments, and \emph{Probably Correct Reliability (PCR)}, where we assign a confidence that the reliability target will be met in a relatively static environment.
\end{enumerate} 

We take a closer look into the impact that the channel uncertainty has on the reliability that can be attained in wireless systems operating in UR-relevant regime. This calls naturally for the use of the statistical learning methodology, which has recently started to get a traction in the wireless communication community~\cite{simeone2018very}.  
Our investigation, which to the best of our knowledge is the first of its kind, shows that,
the knowledge about the true channel statistics affects significantly the amount of effort that has to be invested in guaranteeing high reliability. Furthermore, compared to the case in which the channel statistics is perfectly known, the transmitter needs to sacrifice the spectral efficiency in order to meet the reliability requirements.
Hence, more robust methods for design and assessment of URLLC systems are required.
We illustrate this in the following example.




\subsection{{A Motivating Example}}
Alice is selling an ultra-reliable wireless communication system.
In the advertisement she claims that the system can deliver information at a rate of up to $R$ with the probability of failing being no larger than $\epsilon$ (e.g. $10^{-5}$).
Bob considers purchasing the system and wants the following two questions answered:
\begin{enumerate}
\item \textit{How does Alice measure the reliability performance?}
\item \textit{Under {what conditions can the system offer the advertised performance?}}
\end{enumerate}
To investigate, Bob contacts Alice {and gets the following response}: \textit{when established over a flat fading wireless channel with received power {drawn from a cumulative distribution function} $F$, the system can {support one-way transmission at a maximal} rate $R = R_{\epsilon}(F)$ while maintaining link outage probability equal to $\epsilon$.}
Alice {also provides Bob a list of rates $R_{\epsilon}(F)$ for various values of $F$ and $\epsilon$}.

Alice's answer might sound reassuring to some customers, but not to Bob.
Instead, he finds the answer limiting and responds with several other questions:
\begin{itemize}
\item \textit{What happens if the system has only limited/imperfect knowledge of $F$?}
\item \textit{What if the true channel differs from $F$?}
\item \textit{What if the system has no knowledge of $F$ at all?}
\end{itemize}
Is Alice in a position to make an informed statement about the achievable rates and the corresponding reliability performance of the system under all of the above circumstances?
Although she might be able to give some insights on part of them, Alice has a hard time in giving Bob technically precise or fully general answers to all questions.
This is because, unless $F$ is fully and perfectly known, it is unclear to Alice how the reliability should be assessed. 


\subsection{URLLC Revisited}

This paper answers Bob's questions in a novel \emph{statistical framework} for design and performance assessment. 
Being inspired by supervised learning, the framework {consists of} three key elements: \emph{model selection}, \emph{learning}, and \emph{rate selection}. 

\subsubsection*{Model Selection}
As indicated in the above example, when transmitting at rate $R_{\epsilon}(F)$, the corresponding reliability, assessed in terms of e.g. link outage probability $\epsilon$, can be guaranteed with \emph{certainty} only in the case when the system has perfect knowledge of the true distribution $F$ i.e. the exact channel statistics.
Such knowledge is seldom available in practice and in all other cases the system has to learn, i.e., estimate $F$.
In order to do so, the system first selects a model for $F$.
The choice will, in general, depend on the state of knowledge of the system regarding the true distribution $F$.
In some cases, a side information on the physical properties of the wireless channel might be available.
For instance, the system might know that the channel exhibits a strong diffuse/specular component and adopt the Rayleigh/Rician fading model \cite{vaughan2003channels}. 
In other cases, physical knowledge will be unavailable; hence, no parametric model is suitable, and the system should resort to non-parametric models. 
Alternatively, given that in URLLC applications only the regions of very low outages are of practical interest, the system might resort to simple and general first order approximations of the lower tail of $F$ \cite{Eggers2017}.


\subsubsection*{Learning}
After selecting an appropriate model, the system applies a {learning} procedure that generates an {estimate} of $F$, denoted by $\widehat{F}$, using a finite number of channel measurements.
We refer to the measurements as the \emph{training sample} and assume that they are collected in a dedicated \emph{training phase} prior to transmission.\footnote{The reader will note that the definition of the training phase is vague, i.e., we intentionally do not specify the details on how the channel training is actually performed. In other words, our analysis is valid for variety of channel training schemes from conventional, dedicated pilot signals and training sequences, to previous transmissions, where the ``training'' consists of actual data transmissions in which feedback about the instantaneous channel quality is received.}
For parametric models including the models based on tail approximations, we use \emph{maximum likelihood estimation (MLE)}, as a standard learning tool in absence of informative priors.

\subsubsection*{Rate Selection}
The estimated distribution $\widehat{F}$ is a function of the random training sample which makes it a random quantity itself and is, therefore, inherently {uncertain}.
Choosing the transmission rate as $R=R_{\epsilon}(\widehat{F})$ randomizes the outage probability, i.e., different realizations of the training sample induce different outage probabilities; this is why Alice {is} unable to give Bob deterministic reliability guarantees.
{In other words, the commonly accepted methodology of specifying ultra-reliability through an outage probability value is essentially insufficient since, under limited channel knowledge and uncertainty, the outage probability is a random variable.}
In our framework, the system selects the transmission rate such that predefined {statistical reliability constraints} are satisfied; hence, the reliability now is guaranteed either on average or {probabilistically}. 
These  reliability constraints rely on the statistical characterization of the outage probability as a random variable and impose limits on some specific probabilistic measures.
We consider two {types of} constraints. 
The first type of constraint, termed \emph{Averaged Reliability (AR)}, controls the mean of the outage probability over {all} possible realizations of the training sample and is suitable for designing URLLC systems that perform as desired {on the average}, over all transmissions. {The second constraint}, termed \emph{Probably Correct Reliability (PCR)}, controls the probability $\xi$ that the outage probability violates $\epsilon$ for a given {specific} training sample. {The latter constraint} is more restrictive and suitable for applications that require tighter, {per-transmission} control of the performance of the system.
{In principle, the second constraint generalizes the first as it provides the system designer with the freedom to control the {higher order moments} of the distribution of the outage probability via an additional free parameter $\xi$. 
This type of statistical approach in characterizing the reliability performance probabilistically via {two} parameters is reminiscent to the framework of \emph{probably approximately correct (PAC) learning} \cite{Valiant:1984:TL:1968.1972,shalev2014understanding}. The reader would recall that the goal in PAC learning, after acquiring the training samples, is to select a generalization function which, with {high probability}, has a {low generalization error}. Drawing the parallel, in our framework, when designing the URLLC system according to the second constraint, the goal of the designer, after training the channel, is to select a transmission rate which with high probability $1-\xi$ will have link outage probability equal to $\epsilon$.
Clearly, in both frameworks the performance is assessed {probabilistically}.}


\subsubsection*{Consistency and Reliability Assessment}
As the size of the training sample grows, we intuitively expect that the transmission rate will converge to Alice's rate, i.e., the maximum achievable rate $R_{\epsilon}(F)$ for given $\epsilon$ when $F$ is perfectly known; we refer to this desired property as \emph{consistency}.
We observe that ensuring consistency while meeting the statistical reliability requirements is crucially dependent on the relation between the assumed model and the true distribution.

Summarizing the main findings, we conclude that the scheme is consistent when
\begin{itemize}
\item The assumed model belongs in the same parametric family as $F$, or
\item The system uses a non-parametric model.
\end{itemize}
In the first case, the transmission rate converges rather quickly. {However,} the assumption of knowing which parametric family of models the channel belongs to is a strong one, {difficult to fulfill} in practice and prone to modeling mismatch which severely violates the reliability performance. 
Non-parametric models have generalization power and work for any channel. Nevertheless, the convergence is slow, requiring prohibitive training sample sizes.
Finally, models based on first-order lower tail approximation offer ``the best of the two worlds'' \cite{Eggers2017}.
Although such models do not maintain consistency due to inherent mismatch stemming from the approximation error, they still show superior performance compared to poor parametric modeling choices. In addition, they also require less channel training compared to non-parametric models.

A final remark is in order.
Despite the fact that our work is limited to the simple case of one-way transmission, we note that the statistical treatment we introduce here opens up a methodology that can be readily extended to more advanced and complex transmission and networking scenarios that also introduce other sources of uncertainties, such as multiple antenna techniques, ARQ, interference, etc.

The rest of the paper is organized as follows.
Section~\ref{sec:prelim} introduces the system model.
Section~\ref{sec:new_reliability} introduces novel statistical reliability constraints and formulates the problem.
Sections~\ref{sec:parametric},~\ref{sec:nonparametric} and ~\ref{sec:powerlaw} focus on parametric, non-parametric and approximate channel models, respectively, and derive the corresponding rate-selection functions.
Section~\ref{sec:results} presents and discusses the numerical evaluations and Section~\ref{sec:conc} concludes the paper.

\section{Preliminaries}
\label{sec:prelim}

\subsection{System model}

We consider an one-way communication link where the transmitter (Tx) sends a packet to a receiver (Rx) at rate $R$ over a flat fading wireless channel. 
With $F$ we denote the cumulative distribution function (CDF) of the received power $Y\geq 0$; we denote $Y\sim F$ and use the terms true channel statistics and true distribution interchangeably when referring to $F$.
We assume that $F$ belongs to a class $\mathcal{F}$ of smooth distributions defined over the non-negative reals.
We will consider two cases: 1) $\mathcal{F}$ is a \emph{parametric}, and 2) $\mathcal{F}$ is an arbitrary non-parametric family of distributions.
Prior to transmission, in the {training phase} the Tx collects $n$ independent and identically distributed (i.i.d.) channel measurements from $F$; we refer to them as the \emph{training sample} and denote them by $X^n = \{X_1,\hdots,X_n\}$.\footnote{Throughout the paper, we use small case letters to denote specific realizations of random variables.}

\subsubsection*{Parametric Channel Models}
In this case, we assume that $F$ belongs to a parametric family of distributions  $\mathcal{F}_{\boldsymbol{\theta}}=\{F_{\boldsymbol{\theta}}:\boldsymbol{\theta}\in\mathbf{\Theta}\}$ with $\mathbf{\Theta}$ denoting the parameter space.
The parameter $\boldsymbol{\theta}$, which can be a vector or a  scalar, usually has a specific interpretation stemming from the physical properties of the wireless channel. 
We review three common parametric models that {are widely used} in practice.



\emph{Rayleigh channel:}
Adopted in many wireless studies, the Rayleigh model represents a single scatterer (i.e. cluster) where the received signal is due to a diffuse component only \cite{vaughan2003channels,Durgin2002}. 
The received power $Y$ follows an exponential distribution and, hence, the CDF is given by
\begin{align}\label{eq:cdf_X_rayl}
F_{\lambda}(y) = 1 - e^{-\frac{y}{\lambda}},
\end{align}
where $\lambda = \mathbb{E}\left[Y\right]$ denotes the average received power.

\emph{Rician channel:}
The Rician channel is an extension of the Rayleigh fading model, featuring an additional specular component \cite{vaughan2003channels,Durgin2002}.
Let the power of the specular component be denoted by $\rho$, whereas, similarly to the Rayleigh case, the mean power of the diffuse component is denoted by $\lambda$.
Let $k=\rho/\lambda\geq 0$ be the ratio between the average powers of the specular and diffuse components, also known as a Rician $k$-factor.
The CDF of the received power under Rician fading is given as
\begin{align}\label{eq:cdf_X_rice}
F_{\lambda,k}(y) = 1 - Q_1\lefto(\sqrt{2k},\sqrt{2\frac{y}{\lambda}}\right),
\end{align}
where $Q_1(\cdot, \cdot)$ is the $1^{\text{st}}$ order Marcum Q-function.
The average received power is $\mathbb{E}\left[Y\right] = \rho + \lambda$.
The special case when $k=0$, indicating no specular component, corresponds to Rayleigh-distributed channel as in \eqref{eq:cdf_X_rayl}.


\emph{Nakagami-$m$ channel:}
The extension of the Rayleigh model in multi-cluster settings is the Nakagami-$m$ model where the received envelope follows the Nakagami distribution with shape parameter $m$ and scale parameter $\lambda$ \cite{NAKAGAMI19603}.
The Nakagami-$m$ channel can be interpreted as the incoherent sum of $m$ i.i.d. Rayleigh-type clusters, each with mean diffuse power $\lambda$.
The CDF of the received power under Nakagami-$m$ fading is
\begin{align}\label{eq:cdf_X_Nak}
F_{\lambda,m}(y) = \frac{\gamma\lefto(m,\frac{y}{\lambda}\right)}{\Gamma(m)}
\end{align}
with $\gamma(\cdot,\cdot)$ denoting the lower incomplete gamma function, whereas $\Gamma(\cdot)$ denotes the gamma function.
Evidently, the average received power is $\mathbb{E}\left[Y\right]=m\lambda$.
For generality, we assume that $m\geq 0.5$ \cite{NAKAGAMI19603}.
The special case $m=1$ indicates a single diffuse cluster and therefore corresponds to the Rayleigh channel \eqref{eq:cdf_X_rayl}.


\subsection{Outage probability}
To isolate and study the impact of channel uncertainty, we neglect the impact of noise and interference and consider errors due to \emph{link outage} only; link outages are defined by the following event:
\begin{align}\label{eq:outage_event}
R>\log_2(1+Y).
\end{align}
Hence, the {outage probability} at transmission rate $R$ is defined as
\begin{align}\label{eq:outage_prob}
p_{F}(R) = \mathbb{P}\lefto[R>\log_2(1+Y)\right]. 
\end{align}
The goal of {ultra-reliable communication} is to choose the {maximal rate} that meets a predetermined {reliability constraint}, such as
\begin{align}\label{eq:urc}
p_F(R)\leq\epsilon.
\end{align}
However, designing the reliability criteria as well as determining the most favorable transmission rate is strongly linked to the amount of knowledge, i.e., {state of knowledge} the Tx has about the true distribution. 
{As illustrated in the following section, specifying the reliability performance using only \eqref{eq:urc} when the Tx has limited knowledge of the channel is {no longer sufficient}.}

\section{Reliability Guarantees Under Limited Channel Knowledge}
\label{sec:new_reliability}

\subsection{Perfect channel knowledge: $\epsilon$-outage capacity}

First, consider the benchmark case when the Tx \emph{perfectly} knows $F$.
In such circumstances, the Tx can easily determine the maximum rate as a function of $F$ at which the outage probability is no larger than $\epsilon$, i.e., \eqref{eq:urc} can be guaranteed deterministically:
\begin{align}\label{eq:e_outage_capacity1}
R_{\epsilon}(F) & = \sup\lefto\{R\geq 0:p_{F}(R)\leq\epsilon\right\}\\\label{eq:e_outage_capacity2}
& = \log_2\left(1 + F^{-1}(\epsilon)\right).
\end{align}
The term $R_{\epsilon}(F)$ is also known as the $\epsilon$-outage capacity, whereas $F^{-1}(\epsilon)$ is the $\epsilon$-quantile of $F$.
For parametric models, knowing the channel implies that $\boldsymbol{\theta}$ is known perfectly and we use the notation $R_{\epsilon}(\boldsymbol{\theta})$.

\subsection{Limited channel knowledge: MLE of the $\epsilon$-outage capacity}

To illustrate the impact of channel uncertainty, consider the following: the Tx knows that $Y\sim F_{\boldsymbol{\theta}}$ but has no knowledge of $\boldsymbol{\theta}$.
Having acquired a training sample $x^n$ prior to transmission, the Tx can {learn} $\boldsymbol{\theta}$ via MLE as follows
\begin{align}\label{eq:ML_gen}
\hat{\boldsymbol{\theta}}(x^n) = \text{arg}\max_{\boldsymbol{\theta}\in\mathbf{\Theta}}\sum_{i=1}^n\log F_{\boldsymbol{\theta}}^{'}(x)|_{x=x_i}.
\end{align}
Then, the ``plug-in'' estimator
\begin{align}\label{eq:plugandplay_rate_selection_function}
R(x^n) = R_{\epsilon}\lefto(\hat{\boldsymbol{\theta}}(x^n)\right),
\end{align}
is MLE of the $\epsilon$-outage capacity.
Selecting the rate this way is na{\"i}ve due to the {uncertainty} of $R(X^n)$, which is itself a {random variable}.
It follows from \eqref{eq:outage_prob} that different $R$ yields different outage probability for fixed $F$; in other words, the random sequence $X^n$ induces a distribution on the outage probability and the Tx can no longer guarantee with certainty that the outage probability for the transmission rate $R(X^n)$, denoted by  $p_{\boldsymbol{\theta}}(R(X^n))$, will be less than or equal to $\epsilon$.
In fact, if the probability density function of the outage probability is symmetric (i.e., the mean and the median coincide), 
we have that $\mathbb{P}\left[p_{\boldsymbol{\theta}}(R(X^n))>\epsilon\right]=0.5$ which in some practical setups, as discussed below, is clearly unacceptable. 


\subsection{Problem formulation}

The discussion in the previous subsection shows that, when the Tx has a limited knowledge about the channel, it can only guarantee the reliability in a probabilistic manner.
Formally stated, this is done by {defining} a \emph{rate-selection function} $R(X^n)$ such that a predetermined statistical reliability constraint is satisfied.

\subsubsection{Statistical reliability constraints}

We consider two different approaches, resulting in two types of constraints, each of them suited to a specific set of use cases.

\emph{Averaged Reliability (AR)}:
We consider the probability
\begin{align}\label{eq:outage_prob_appI}
\overline{p}_F = \mathbb{P} \left[R(X^n)>\log_2(1+Y)\right],
\end{align}
computed w.r.t. the joint distribution of $X^n$ and $Y$.
Using the law of total expectation, \eqref{eq:outage_prob_appI} can be rewritten as
\begin{align}
\overline{p}_F & = \mathbb{E}\lefto[\mathbb{P}\lefto[R(X^n)>\log_2(1+Y)|X^n\right]\right]\\
& = \mathbb{E}\lefto[p_F(R(X^n))\right],
\end{align}
where the outer expectation is taken over the distribution of the training sample $X^n$.
It follows that $\overline{p}_F$ is the \emph{mean} of the outage probability, as defined in \eqref{eq:outage_prob}, averaged over the training sample $X^n$.

We consider the reliability constraint
\begin{align}\label{eq:constI}
\sup_{F\in\mathcal{F}} \overline{p}_F\leq\epsilon
\end{align}
which controls the \emph{worst-case} mean outage probability over the whole class $\mathcal{F}$ and provides a firm reliability guarantee. 

{The mean outage probability \eqref{eq:outage_prob_appI} can be used when the Tx's objective is to optimize the transmission rate \emph{jointly} over the training and the transmission, when channel training is performed \emph{prior} to each transmission.}
This approach is suitable in a dynamic environment in which the channel changes frequently, requiring frequent channel training and estimation; an example of this is a vehicular communication scenario.



\emph{Probably Correct Reliability (PCR)}:
Differently from the cases suited for AR, in many URLLC applications, such as monitoring and control in power grids or automated production in industrial complexes, the environment remains reasonably static for long periods of time. In such cases, it makes sense to train the channel infrequently implying that the most recent channel estimate will be used by the system over many future transmissions.
Clearly, the Tx has to be more conservative here and choose the rate such that the outage probability in all the following transmissions is below $\epsilon$ with high probability. 

This scenario is captured by PCR, which is more restrictive and effectively controls the higher order moments of the distribution of the outage probability. PCR is suitable when the Tx sets the transmission rate for all future transmissions \emph{after} obtaining the training sample.
We rely on the concept of meta-probability \cite{Haenggi2016} and introduce:
\begin{align}\label{eq:outage_prob_appII}
\widetilde{p}_F & =  \mathbb{P}\left[\mathbb{P}\lefto[R(X^n)>\log_2(1+Y)|X^n\right]>\epsilon\right]\\
& = \mathbb{P}\lefto[p_F(R(X^n))>\epsilon\right],
\end{align}
where the outer probability is computed w.r.t. the joint distribution of the training sample $X^n$.
In PCR we define the statistical reliability constraint
\begin{align}\label{eq:constII}
\sup_{F\in\mathcal{F}} \widetilde{p}_F \leq\xi,
\end{align}
over the class $\mathcal{F}$.
The probability $\xi$ bounds the worst-case probability that the conditional outage probability \eqref{eq:outage_prob} given that $X^n$ is larger than $\epsilon$. Borrowing the terminology from PAC learning, $\xi$ is the confidence parameter that indicates how likely it is to meet the reliability requirement \cite{shalev2014understanding}.


\subsubsection{Rate-selection function}

There is a whole family of rate-selection functions $R(X^n)$ that satisfy \eqref{eq:constI} or \eqref{eq:constII}. 
In order to find the most favorable, one should introduce an objective function of $R(X^n)$ and define an optimization problem that will give the \emph{optimal} rate-selection function subject to \eqref{eq:constI}/\eqref{eq:constII}.
A possible objective function might be the ratio
\begin{align}\label{eq:average_th_ratio_def}
\omega_{\epsilon}^n(F) = \frac{\mathbb{E}\mathopen{}\left[R(X^n)1_{R(X^n)\leq\log_2(1+Y)}\right]}{R_{\epsilon}(F)(1-\epsilon)},
\end{align}
between the throughput using $R(X^n)$ and the optimal throughput, given that $F$ is known perfectly.
Formulating such optimization problem using \eqref{eq:average_th_ratio_def} is a non-trivial task and therefore out of the scope of the paper.
One issue is the fact that \eqref{eq:average_th_ratio_def} depends on a specific and fixed true distribution $F$ while our aim is to design a robust rate-selection function that maximizes the transmission rate over potentially large class $\mathcal{F}$ of channel distributions.

In the rest of the paper, we will limit our discussion to the heuristic, yet intuitive choice, inspired by the na{\"i}ve, MLE-based approach described in the previous subsection.
Namely, the Tx uses
\begin{align}\label{eq:rate_selection_function}
R(X^n) = \log_2\lefto(1 + \widehat{F}^{-1}(\varepsilon_n)\right), 
\end{align}
where $\widehat{F}^{-1}(\varepsilon_n)$ is an estimate of the $\varepsilon_n$-quantile of the channel for some positive sequence $\varepsilon_n>0$ and training sample $X^n$.
The objective now is to find $\varepsilon_n$ that \emph{maximizes} $R(X^n)$ while meeting either \eqref{eq:constI} or \eqref{eq:constII}.
Note that if $\varepsilon_n = \epsilon$ for every $n$, we have the  plug-in solution where $R(X^n)$ is just the MLE of the $\epsilon$-outage capacity.
Note that the rate-selection function \eqref{eq:rate_selection_function} is still an estimate of the $\epsilon$-outage capacity; however, by choosing $\varepsilon_n$ such that \eqref{eq:constI}/\eqref{eq:constII} is satisfied, intuitively the Tx controls the uncertainty of the transmission rate introduced by limited channel knowledge.

We next introduce the notion of \emph{consistency}. Namely, a rate-selection function is said to be consistent if $R(X^n)$ converges to the $\epsilon$-outage capacity $R_{\epsilon}(F)$  as $n\rightarrow\infty$ with probability $1$ for all $F\in\mathcal{F}$ while simultaneously satisfying either \eqref{eq:constI} or \eqref{eq:constII}.
In such case
\begin{align}
\lim_{n\rightarrow\infty}\omega_{\epsilon}^n(F) = 1.
\end{align}
Clearly, consistent rate-selection functions are desirable, but,  depending on the relation between the model and the true distribution, not always possible. 
Specifically, if there is a model mismatch, such that the 
model differs from the actual channel distribution, 
then the rate-selection function will not be consistent and $\lim_{n\rightarrow\infty}\omega_{\epsilon}^n(F) \neq 1$. This phenomenon, which can be also linked to the {bias-variance trade-off} \cite{shalev2014understanding}, is discussed in more detail in  the following section.



\section{Parametric Rate-selection Functions}\label{sec:parametric}

We begin by considering parametric channel models.
In this case, \eqref{eq:rate_selection_function} can be rewritten as
\begin{align}\label{eq:rate_selection_function_param}
R(X^n) = R_{\varepsilon_n}\lefto(\hat{\boldsymbol{\theta}}(X^n)\right)
\end{align}
for some $\varepsilon_n>0$; here, $\hat{\boldsymbol{\theta}}$ is the MLE of the parameter $\boldsymbol{\theta}$ using the training sample $X^n$.
We illustrate some of our main insights through several case studies for which we impose different assumptions regarding the true distribution and the state of knowledge at the Tx. 
In particular, we focus on the following cases:
\begin{enumerate}
\item The channel is Rayleigh-distributed and the Tx knows this; however, the Tx has no knowledge of $\lambda$.
\item The channel is {not} Rayleigh distributed and the Tx does not know this; {nevertheless, the Tx still assumes that the channel is Rayleigh-distributed, with unknown $\lambda$}.
\end{enumerate}
For known $\lambda$, the $\epsilon$-outage capacity is easily computed as
\begin{align}\label{eq:e_outage_capacity_rayleigh}
R_{\epsilon}(\lambda) = \log_2\lefto(1 - \lambda\log(1-\epsilon)\right).
\end{align}
Note that, even for the simple and intuitive choice \eqref{eq:rate_selection_function_param}, finding $\varepsilon_n$ that maximizes $R(X^n)$ such that \eqref{eq:constI}/\eqref{eq:constII} is satisfied, is a non-trivial exercise for most of the remaining parametric channel models and one needs to resort to numerical methods to compute the transmission rate. 


\subsection{The true distribution is Rayleigh}

The MLE of $\lambda$ under Rayleigh fading is just the sample mean and the transmission rate for specific training sample $x^n$ becomes
\begin{align}\label{eq:rate_selection_function_rayleigh}
R(x^n) = \log_2\lefto(1 - \frac{\log(1-\varepsilon_n)}{n}\sum_{i=1}^n x_i \right).
\end{align}
We show how to find $\varepsilon_n$ for any $n$ such that \eqref{eq:rate_selection_function_rayleigh} is maximized and either \eqref{eq:constI} or \eqref{eq:constII} is satisfied.
Specifically, the AR as defined in \eqref{eq:outage_prob_appI} can be computed as
\begin{align}\label{eq:casestudy1_constI_1}
\overline{p}_{\lambda} & = 
\mathbb{E}\lefto[F_{\lambda}\lefto(- \frac{\log(1-\varepsilon_n)}{n}\sum_{i=1}^n X_i\right)\right]\\\label{eq:casestudy1_constI_3}
& = 1 - \mathbb{E}\lefto[\exp\lefto\{\frac{\log(1-\varepsilon_n)}{n\lambda}\sum_{i=1}^n X_i\right\}\right]\\\label{eq:casestudy1_constI_4}
& = 1 - \left(1 - \frac{\log(1-\varepsilon_n)}{n}\right)^{-n}.
\end{align}
In \eqref{eq:casestudy1_constI_3} we have used \eqref{eq:cdf_X_rayl} and in \eqref{eq:casestudy1_constI_4} the moment generating function (MGF) of an exponential random variable with mean $\lambda$ given by $M_{\lambda}(t) = (1 - t\lambda)^{-1}$.
Now, $\varepsilon_n$ can be computed from \eqref{eq:constI} by equating \eqref{eq:casestudy1_constI_4} with $\epsilon$; we obtain
\begin{align}\label{eq:varen_constI_rayleigh}
\varepsilon_{n} = 1 - e^{-n\left((1-\epsilon)^{-\frac{1}{n}} - 1\right)}.
\end{align}
Interestingly, for the specific case of Rayleigh channel, the AR given by \eqref{eq:varen_constI_rayleigh} does not depend on $\lambda$ and selecting $\varepsilon_n$ according to \eqref{eq:varen_constI_rayleigh} gives a transmission rate that satisfies \eqref{eq:constI} for all $\lambda$.

Similarly, the meta-probability \eqref{eq:outage_prob_appII} in PCR evaluates to
\begin{align}\label{eq:casestudy1_constII_1}
\widetilde{p}_{\lambda} & = 
\mathbb{P}\lefto[F_{\lambda}\lefto(- \frac{\log(1-\varepsilon_n)}{n}\sum_{i=1}^n X_i\right)>\epsilon\right]\\\label{eq:casestudy1_constII_3}
& = \mathbb{P}\lefto[1 - \exp\lefto\{\frac{\log(1-\varepsilon_n)}{n\lambda}\sum_{i=1}^n X_i\right\}>\epsilon\right]\\\label{eq:casestudy1_constII_4}
& = \mathbb{P}\lefto[\sum_{i=1}^nX_i>n\lambda\frac{\log(1-\epsilon)}{\log(1-\varepsilon_n)}\right]\\
\label{eq:casestudy1_constII_5}
& = 1 - \frac{\gamma\left(n,n\frac{\log(1-\epsilon)}{\log(1-\varepsilon_n)}\right)}{(n-1)!}.
\end{align}
In \eqref{eq:casestudy1_constII_5} we used the CDF of an Erlang-$n$ random variable with shape parameters $\lambda$ (obtained as a sum of $n$ i.i.d. exponential random variables with mean $\lambda$), given by $\mathbb{P}\left[\sum_{i=1}^n X \leq x\right]=\gamma(n,x\lambda)/(n-1)!$. 
Observe again that \eqref{eq:casestudy1_constII_5} does not depend on $\lambda$.
By choosing  $\varepsilon_n$ as the maximum value satisfying
\begin{align}\label{eq:varen_constII_rayleigh}
1 - \frac{\gamma\left(n,n\frac{\log(1-\epsilon)}{\log(1-\varepsilon_n)}\right)}{(n-1)!} \leq \xi,
\end{align}
we obtain a rate-selection function that meets \eqref{eq:constII} for any $\lambda$.



\subsection{Mismatch: The true distribution is not Rayleigh}\label{sec:mismatched}

Next, we study the impact of channel mismatch on the reliability performance of the system.
We assume that the channel is no longer Rayleigh, i.e., $F_{\boldsymbol{\theta}}$ is different from \eqref{eq:cdf_X_rayl}; yet, the Tx maintains the assumption that the channel is Rayleigh, setting the rate as in \eqref{eq:rate_selection_function_rayleigh} with $\varepsilon_n$ computed via \eqref{eq:varen_constI_rayleigh}/\eqref{eq:varen_constII_rayleigh}.
Due to mismatch, the rate-selection function can no longer be guaranteed to be consistent, i.e., $R(X^n)$ does not converge to $R_{\epsilon}(\boldsymbol{\theta})$ computed w.r.t. the true distribution $F_{\boldsymbol{\theta}}$.
As a result, one can no longer guarantee that the reliability constraints \eqref{eq:constI} or \eqref{eq:constII} will be satisfied.

Given that the Tx sets the transmission rate as in \eqref{eq:rate_selection_function_rayleigh}, the mean outage probability \eqref{eq:outage_prob_appI} and the meta-probability \eqref{eq:outage_prob_appII} can be written as
\begin{align}\label{eq:mean_outage_miss}
\overline{p}_{\boldsymbol{\theta}} & = \mathbb{E}\lefto[F_{\boldsymbol{\theta}}\lefto(-\frac{\log(1-\varepsilon_n)}{n}\sum_{i=1}^n X_i\right)\right],\\\label{eq:metaprob_miss}
\tilde{p}_{\boldsymbol{\theta}} & = \mathbb{P}\lefto[F_{\boldsymbol{\theta}}\lefto(-\frac{\log(1-\varepsilon_n)}{n}\sum_{i=1}^n X_i\right)>\epsilon\right],
\end{align}
where the outer expectations are taken w.r.t. the true distribution $F_{\boldsymbol{\theta}}$.
The above quantities can be easily evaluated numerically for any $F_{\boldsymbol{\theta}}$.
To gain more insight into the impact of channel mismatch, we derive simple approximations relaying on the assumption that the outage probability conditioned on $x^n$ is small, i.e., in the order of $\epsilon$.
In such case, for wide variety of channels, $F_{\boldsymbol{\theta}}$ can be approximated via simple \emph{power law} (see \cite{Eggers2017} for detailed derivations based on first-order approximations for different channels and Section~\ref{sec:powerlaw} for an argument from the extreme value theory)
\begin{align}\label{eq:cdf_powerlaw}
F_{\boldsymbol{\theta}}(y)\approx \alpha_{\boldsymbol{\theta}}y^{1/\kappa_{\boldsymbol{\theta}}},\;y\rightarrow 0,
\end{align}
where $\alpha_{\boldsymbol{\theta}}$ and $\kappa_{\boldsymbol{\theta}}$ depend on the true distribution $F_{\boldsymbol{\theta}}$.
Then, we obtain the simple approximations
\begin{align}\label{eq:mismatch_appI_1}
\overline{p}_{\boldsymbol{\theta}}
& \approx \alpha_{\boldsymbol{\theta}}\left(-\log(1-\varepsilon_n){\mathbb{E}}_{\boldsymbol{\theta}}\left[X\right]\right)^{1/\kappa_{\boldsymbol{\theta}}}\left( 1 + \frac{1-\kappa_{\boldsymbol{\theta}}}{2 n \kappa_{\boldsymbol{\theta}}^2}\frac{\text{Var}_{\boldsymbol{\theta}}[X]}{\left({\mathbb{E}}_{\boldsymbol{\theta}}\left[X\right]\right)^{2}}\right),\\
\widetilde{p}_{\boldsymbol{\theta}} 
\label{eq:mismarch_appII_1}
& \approx \exp{\left\{\frac{t^*n}{\log(1-\varepsilon_n)}\left(\frac{\epsilon}{\alpha_{\boldsymbol{\theta}}}\right)^{\kappa_{\boldsymbol{\theta}}}\right\}}(M_{\boldsymbol{\theta}}(t^*))^n.
\end{align}
In \eqref{eq:mismatch_appI_1} we used a second-order Taylor expansion to approximate the expectation $\mathbb{E}\left[\left(\sum_i X_i\right)^{1/\kappa_{\boldsymbol{\theta}}}\right]$ and in \eqref{eq:mismarch_appII_1} we used the Chernoff method to approximate $\mathbb{P}\left[\sum_iX_i>x\right]$; $M_{\boldsymbol{\theta}}(t^*)$ is MGF of $X$ with $t^*$ obtained as a solution to the minimization problem
\begin{align}
t^* = \inf_{t>0}\exp{\left\{\frac{tn}{\log(1-\varepsilon_n)}\left(\frac{\epsilon}{\alpha_{\boldsymbol{\theta}}}\right)^{\kappa_{\boldsymbol{\theta}}}\right\}}(M_{\boldsymbol{\theta}}(t))^n.
\end{align}

\emph{Examples:}
Consider the Ricean channel \eqref{eq:cdf_X_rice} whose power law approximation has the form \cite{Eggers2017}
\begin{align}
F_{k,\lambda}(y) \approx \frac{e^{-k}}{\lambda}y,
\end{align}
i.e., $\alpha_{k,\lambda} =  e^{-k}/\lambda$ and $\kappa_{k,\lambda}=1$.
Note that the Rician channel asymptotically exhibits the same slope as the Rayleigh channel but with different scaling.  
The mean outage probability is
\begin{align}\label{eq:mismatch_rice_appI_2}
\overline{p}_{\lambda,k} \approx - \frac{k+1}{e^{k}}\log(1-\varepsilon_n).
\end{align}
Note that under Rayleigh fading, $k=0$ and \eqref{eq:casestudy1_constI_4} can be approximated as $\overline{p}_{\lambda}\approx - \log(1-\varepsilon_n)$. 
Hence, we see from \eqref{eq:mismatch_rice_appI_2} that in case of Rician fading but with transmission rate dimensioned for the empirical channel mean, the mean outage probability is always smaller than the target $\epsilon$, i.e., the transmission rate will always be pessimistic and lower than the maximum rate the Rician channel can support for outage $\epsilon$.

Now, let us consider the Nakagami-$m$ channel for $0.5\leq m\leq1$ (note that for $m=0.5$, the distribution of received amplitude $\sqrt{Y}$ under Nakagami-$m$ fading is exponential); the power law approximation obtains the form
\begin{align}\label{eq:nakagami_powerlaw}
F_{\lambda,m}(y) \approx \frac{1}{\lambda^m \Gamma(m+1)}y^m
\end{align}
and the mean outage probability can be approximated as
\begin{align}\label{eq:mismatch_nakagami_appI_3}
\overline{p}_{\lambda,m} \approx \frac{(-m)^m}{\Gamma(m+1)}(\log(1-\varepsilon_n))^m.
\end{align}
It is easy to check that for $0.5\leq m\leq 1$, the above expression is always larger than $\epsilon$, i.e., the transmission rate will be always optimistic and larger than the maximum rate the Nakagami-$m$ channel offers for given $\epsilon$.
These results fit well the intuition: for equal average power $\lambda$ of the diffuse component, the Rayleigh CDF \eqref{eq:cdf_X_rayl} is always an upper/lower bound bound on the Rician/Nakagami-$m$ (for $m < 1$) CDFs given in \eqref{eq:cdf_X_rice}/\eqref{eq:cdf_X_Nak}.

Another interesting observation that can be deduced from \eqref{eq:casestudy1_constI_4}, \eqref{eq:mismatch_rice_appI_2} and \eqref{eq:mismatch_nakagami_appI_3} is the impact of the training sample size $n$ on the convergence.
Provided that $\varepsilon_n$ is selected as in \eqref{eq:varen_constI_rayleigh}, we have that ${-\log(1-\varepsilon_n)} = n\left((1-\epsilon)^{-\frac{1}{n}}-1\right)$.
Given that $\epsilon/n\ll 1$ for any $n\geq 1$, and applying the binomial approximation, we obtain
\begin{align}\label{eq:mean_outage_weak_n}
\overline{p}_{\lambda} \approx \epsilon,\;\overline{p}_{\lambda,k} \approx \frac{k+1}{e^k}\epsilon,\;\overline{p}_{\lambda,m} \approx \frac{m^m}{\Gamma(m+1)}\epsilon^m.
\end{align}
We conclude that the mean outage probability depends very weakly on $n$.
This implies that the transmission rate calculated under AR \eqref{eq:constI} converges quickly.

To compute the PCR, the MGF of $X$ under Rician fading is
\begin{align}\label{eq:mgf_rice}
M_{\lambda,k}(t) = \frac{\exp{\left\{\frac{k\lambda t}{1 - \lambda t}\right\}}}{1-\lambda t},
\end{align}
for $t<1/\lambda$.
Hence, we obtain
\begin{align}
\widetilde{p}_{\lambda,k} \approx \frac{\exp\left\{\left(\frac{k}{1 - \lambda t^*} + \frac{\epsilon e^k}{\log(1 - \varepsilon_n)}\right)n \lambda t^*\right\}}{(1 - \lambda t^*)^n},
\end{align}
and $t^*$ is the solution to the quadratic equation
\begin{align}
n\lambda(1 - \lambda t + k) + \frac{\epsilon n \lambda e^k}{\log(1-\varepsilon_n)}(1-\lambda t)^2 = 0,
\end{align}
that satisfies $t<1/\lambda$.
Similarly, the MGF of $X$ under Nakagami-$m$ fading is
\begin{align}
M_{\lambda,m}(t) = \frac{1}{(1 - t\lambda)^m},
\end{align}
for $t<1/\lambda$; the meta-probability can be approximated as
\begin{align}
\widetilde{p}_{\lambda,m}\approx \frac{\exp\left\{\frac{t^*n\lambda}{\log(1-\varepsilon_n)}(\epsilon\Gamma(m+1))^{\frac{1}{m}}\right\}}{(1 - t^*\lambda)^{mn}}
\end{align}
with 
\begin{align}
t^* = \frac{1}{\lambda}\left(1 + m\log(1-\varepsilon_n)(\epsilon\Gamma(m+1))^{-1/m}\right).
\end{align}
The resulting approximations are less insightful than the approximations for the mean outage probability; in Section~\ref{sec:results} we show numerically that the mismatch impacts the meta-probability performance more severely.



\section{Non-parametric Rate-selection Functions}\label{sec:nonparametric}

Using non-parametric rate-selection approach is suitable when the Tx has no, or very limited knowledge of the channel distribution and is unwilling to impose assumptions that might lead to mismatch and compromise the reliability performance.
Clearly, one major advantage of the non-parametric approach is its generality and versatility, i.e., it is applicable to wide variety of channels, subject only to some smoothness constraints, such as e.g. existence of the first-order derivative of $F$. 
However, this comes at the expense of the duration of the channel training phase; in general, non-parametric approaches often require excessive training sample lengths for reasonably reliable system performance.


\subsection{Non-parametric estimation of the $\epsilon$-outage capacity}

Similarly as in the parametric case, we will first look into the plug-in estimate of the $\epsilon$-outage capacity.
We consider a general case: $Y$ is drawn from an arbitrary distribution $F$, defined over the non-negative reals.
Consider a specific training sample $x^n$; the empirical CDF
\begin{align}\label{eq:ecdf}
\widehat{F}(y) = \frac{1}{n}\sum_{i=1}^n 1_{x_i\leq y}
\end{align}
for $y\geq 0$ serves as an estimate of $F$.
Let $X_{(1)}\leq\hdots\leq X_{(n)}$ denote the order statistics formed from the elements of the $X^n$ and define for convenience $X_{(0)} = 0$ and $X_{(n+1)} = \infty$. Then, for every $y\geq 0$, there exists an integer $i\in\{1,\ldots,n+1\}$ such that
\begin{equation}\label{eq:ecdf2}
\widehat{F}(y) = \frac{i-1}{n} \quad\text{and}\quad x_{(i-1)}\leq y < x_{(i)}.
\end{equation}
Using \eqref{eq:ecdf2}, one can easily derive the plug in estimate of the $\epsilon$-outage capacity as
\begin{align}
R(x^n) & = \sup\lefto\{R>0: \widehat{F}\left(2^R -1\right)\leq\epsilon\right\}\\
& = \sup\lefto\{R>0: \exists i\in\{1,\ldots,n+1\},x_{(i-1)}\leq 2^R - 1 < x_{(i)}, i\leq n\epsilon +1\right\}\\
& = \sup\lefto\{R>0: x_{(\lfloor n\epsilon + 1\rfloor-1)}\leq 2^R - 1 < x_{(\lfloor n\epsilon + 1\rfloor)}\right\}\\\label{eq:outage_capacity_nonparam0}
& = \sup\lefto\{R>0: \log_2(1 + x_{(\lfloor n\epsilon + 1\rfloor-1)})\leq R < \log_2(1 + x_{(\lfloor n\epsilon + 1\rfloor)})\right\}\\\label{eq:outage_capacity_nonparam}
& = \log_2(1 + x_{(\lfloor n\epsilon + 1 \rfloor)}).
\end{align}
From \eqref{eq:outage_capacity_nonparam0} we observe that the number of samples necessary to obtain reliability performance of $\epsilon$ should satisfy $\lfloor n\epsilon +1\rfloor>0$ or, equivalently $n\epsilon>1$; we see that the number of channel samples in purely non-parametric setup grows as
\begin{align}
n\sim\frac{1}{\epsilon}.
\end{align}
In the context of URLLC with target $\epsilon$ in the order of $10^{-6}$ and below, the non-parametric approach requires excessive channel training. 

\subsection{Non-parametric rate-selection function}

We see from \eqref{eq:outage_capacity_nonparam} that in non-parametric setup the $l$-th order statistic $x_{(l)}$ is an estimate of the $\epsilon$-quantile.
Motivated by this, for a given training sample $x^n$, the Tx fixes the rate as
\begin{align}\label{eq:rate_selection_function_nonparam}
R(x^n) = \log_2(1 + x_{(l)})
\end{align}
and chooses the largest $l$ that meets the statistical reliability constraints.
Thus, using \eqref{eq:rate_selection_function_nonparam}, the outage probability \eqref{eq:outage_prob} conditioned on $X^n$ obtains the simple form
\begin{align}
p_{F}(R(X^n)) & = \mathbb{P}\left[R(X^n)\geq\log_2(1+Y)|X^n\right]\\
& = F(X_{(l)}).
\end{align}
Now, let $U_1,\hdots,U_n$ be independent random variables uniformly distributed $[0,1]$.
Then, $F(X_{(l)})$ has the same  distribution as $U_{(l)}$ which is beta-distributed with shape parameters $l$ and $n+1-l$.
With this property, we easily evaluate the mean outage probability \eqref{eq:outage_prob_appI} as
\begin{align}
\overline{p}_F & = \mathbb{E}\left[p_{F}(R(X^n))\right] = \mathbb{E}\left[U_{(l)}\right]\\
& = \frac{l}{n+1}.
\end{align}
Clearly, the mean outage does not depend on $F$ and $\sup_{F\in\mathcal{F}} \overline{p}_F = l/(n+1)$.
Hence, the constraint \eqref{eq:constI} gives
\begin{align}
l \leq \epsilon(n+1).
\end{align}
From the above rate-selection rule, it is evident that when the number of training samples $n<1/\epsilon - 1$ which corresponds to $l<1$, the transmission rate is $R(x^n)=0$.

Similarly, for the meta-probability \eqref{eq:outage_prob_appII}, we obtain
\begin{align}
\tilde{p}_F & = \mathbb{P}\left[p_{F}(R(X^n))>\epsilon\right] = \mathbb{P}\left[U_{(l)}>\epsilon\right]\\
& = 1 - I_{\epsilon}(l,n+1-l),
\end{align}
where $I_x(a,b)$ is the regularized incomplete beta function.
As in the mean outage approach, the meta-probability does not depend on $F$ and the constraint \eqref{eq:constII} implies that $l$ should be chosen as a solution to the equation:
\begin{align}
1 - I_{\epsilon}(l,n+1-l) = \xi.
\end{align}
Since $l$ is an integer, we choose $l$ to be the largest integer satisfying $1 - I_{\epsilon}(l,n+1-l) \leq \xi$.
Albeit the above implicit equation does not reveal immediate insights, the numerical evaluations in Section~\ref{sec:results} show that the meta-probability constraint requires even bigger training samples sizes.

\section{Power law approximation of the channel tail}\label{sec:powerlaw}

In Section~\ref{sec:parametric}, we saw that the training sample size required to learn the channel for parametric channel models is relatively low; however, the model mismatch may have a significant impact on the realized reliability.
The non-parametric method in Section~\ref{sec:nonparametric} does not suffer from this drawback, but requires a training sample size of the order $1/\epsilon$ which is enormous for many practical applications.
This section takes an approximate approach using power law approximations and asymptotic properties.
To this end, we first  argue why the power law approximation for lower tail of $F$, introduced in \eqref{eq:cdf_powerlaw} is of interest.

The Pickands-Balkema-de Haan theorem theorem (see \cite[Th.~2.1.1]{Falk2010} or \cite[Th.~21.17]{Vaart}) in extreme value theory states that, for a large class of distributions $F$, there exists a constant $\kappa > 0$ such that 
\begin{IEEEeqnarray}{rCl} 
\lim_{t\rightarrow 0}\frac{F(ty)}{F(t)} = y^{1/\kappa}\label{eq:pickands}
\end{IEEEeqnarray}
for every  $y > 0$.
Hence, justifying the use of the \emph{power law} approximation
\begin{IEEEeqnarray}{rCl}
    F(y) &\approx& \alpha y^{1/\kappa}\label{eq:power_approximation}
\end{IEEEeqnarray}
for small $y\geq 0$. Here, $\alpha$ and $\kappa$ are parameters that depend on the true but unknown  distribution $F$. For convenience, throughout the section, we  consider the transformed variables $Z = \log(Y)$ and $Z^n=\{Z_i = \log(X_i)\}_{i=1}^n$. After this transformation, the power law approximation in \eqref{eq:power_approximation} implies that
\begin{IEEEeqnarray}{rCl}
F_Z(z) &\approx& \alpha \e{z/\kappa}.\label{eq:power_approximation2}
\end{IEEEeqnarray}
Under the additional assumption that $F$ is smooth and that $F'(y)>0$ for $y>0$, the density of $Z$ is also well-approximated as $f_Z(z) \approx \frac{\alpha}{\kappa}\e{z/\kappa}$.

The importance of the power law approximation is that it allows one to treat the tail of any distribution $F$, satisfying the conditions stated before, as a parametric distribution of only two parameters. We remark that there are three domains of attraction for the extreme value distribution and that \eqref{eq:pickands} only captures one of these, see \cite{Eggers2017} for a list of common fading distributions satisfying the power law approximation and \cite[Th.~2.1.2]{Falk2010} sufficient conditions.
As a result, we can take an approach similar to the one in Section~\ref{sec:parametric} for parametric distributions even though no full parameterization of $F$ is given. 

In order to estimate $\alpha$ and $\kappa$, we shall apply a variation of the ML estimator. In particular, given a training sample $z^n$ and a small constant $\beta$, we use only the $l = \lceil \beta n\rceil$ smallest order statistics $z_{(1)},\ldots,z_{(l)}$ and choose the parameters $\hat \alpha$ and $\hat \kappa$ such that the likelihood $f(z_{(1)},\ldots,z_{(l)}; \hat \alpha, \hat \kappa)$ is maximized. The intuition is that only the smallest observations contain information about the tail of $F$. 

This approximate approach has some clear disadvantages; namely, the range of validity of the power law approximation depends crucially on the value of $\beta$ and on the true distribution $F$. Hence, no strict statistical guarantees can be derived. It is, however, worth noting that in cases where no parameterization of $F$ is available, only the non-parametric approach can provide such true statistical guarantees for reliability at the expense of a large required sample size, and approximate statistical guarantees may therefore be favored.

By using the power law approximation in \eqref{eq:power_approximation} for  approximating $f(z_{(1)},\ldots,z_{(l)}; \hat \alpha, \hat \kappa)$, we find that \cite[Sec.~7.2.4]{Castillo}
\begin{align}
f(z_{(1)},\ldots,z_{(l)}; \alpha, \kappa) & = \frac{n!}{(n-l)!} F_Z(z_{(l)})^{n-l} \prod_{i=1}^l f_Z(z_i) \\
    &\approx  \left( \frac{\alpha}{\kappa}\right)^l\frac{n!}{(n-l)!} (1-\alpha \e{z_{(l)}/\kappa})^{n-l} \e{\frac{1}{\kappa}\sum_{i={1}}^l z_{(i)}}.\label{eq:joint_density}
\end{align}
As a result, by differentiating the logarithm of the right-hand side of  \eqref{eq:joint_density} with respect to $ \alpha$ and $ \kappa$, by equating with zero, and by solving for $ \alpha$ and $ \kappa$, we find the following simple expressions for the estimator $(\hat \alpha,\hat \kappa)$
\begin{IEEEeqnarray}{rCl}\label{eq:kappa_est}
    \hat \kappa &=& \frac{1}{l} \sum_{i=1}^l (Z_{(l)} -Z_{(i)})\\\label{eq:alpha_est}
    \hat \alpha&=& \frac{l}{n} \e{-Z_{(l)}/\hat \kappa}.
\end{IEEEeqnarray}
We let $\widehat F_Z(z)$ be the estimate of the tail distribution of $Z$ given by $\hat \alpha \e{z/\hat \kappa}$. Then, we obtain an estimate of the lower $\varepsilon_n$-quantile of $Z$ as follows
\begin{IEEEeqnarray}{rCl}
    \widehat F_Z^{-1}(\varepsilon_n) 
    &=& \hat\kappa \log\farg{\frac{\varepsilon_n}{\hat\alpha}}\\
    &=& Z_{(l)} + \frac{1}{l}\log\farg{\frac{n\varepsilon_n}{l}}  \sum_{i=1}^l (Z_{(l)}-Z_{(i)}).\IEEEeqnarraynumspace\label{eq:hatFZ}
\end{IEEEeqnarray}
\begin{sloppypar}\noindent 
Under the power law approximation in \eqref{eq:power_approximation2}, one can show that $Z_{(l)}$ and $\sum_{i=1}^l (Z_{(l)}-Z_{(i)})$ are independent random variables. In particular, $Z_{(l)}$ is asymptotically normal and $\sum_{i=1}^l (Z_{(l)}-Z_{(i)})$ is Erlang distributed with parameters $l-1$ and $1/\kappa$ and its distribution is not altered when conditioning on $Z_{(l)}$.
\end{sloppypar}

We now set the rate-selection function as
\begin{IEEEeqnarray}{rCl}\label{eq:rate_selection_powerlaw}
    R(X^n) &=& \log_2\farg{1 + \ee{\widehat F_Z^{-1}(\varepsilon_n)}}
\end{IEEEeqnarray}
where the exponential function is introduced to invert the log-transformation.

We consider two different approaches for specifying $\varepsilon_n$. The first approach relies on asymptotic approximations of $\widehat F_Z^{-1}(\varepsilon_n)$ which holds in the limit $n\rightarrow \infty$. The second approach is simpler and does not rely on asymptotic approximations; however, it is computationally more tedious and does not provide simple analytical insights.

\paragraph*{Asymptotic approach}
Relying on the asymptotic normality of $\widehat F_Z^{-1}(\varepsilon_n)$, we choose $\varepsilon_n$ according to the following simple formula
\begin{IEEEeqnarray}{rCl} 
\varepsilon_n &=& \epsilon\ee{- \sqrt{\frac{\overline V}{n} } Q^{-1}(\xi)}\label{eq:asymp_vareps}
\end{IEEEeqnarray}
where
\begin{IEEEeqnarray}{rCl}
   \overline V =  \frac{1}{\beta} \left(1 - \beta + \log^2 \frac{\epsilon}{\beta}\right).
\end{IEEEeqnarray}
Then, as  shown in Appendix~\ref{app:powerlaw}, we obtain the following approximate meta-probability for PCR:
\begin{IEEEeqnarray}{rCl}
    \lim_{n\rightarrow \infty} \widetilde{p}_F 
    &\approx& \xi\label{eq:meta_prob_asymp}
\end{IEEEeqnarray}
where the approximation stems from the power law approximation in \eqref{eq:power_approximation2}. Hence, the approximation becomes increasingly more accurate as $\beta$ is lowered towards zero. 
Similarly, it can be shown that $\varepsilon_n = \epsilon$ implies that
\begin{IEEEeqnarray}{rCl} 
 \lim_{n\rightarrow \infty} \overline{p}_F
    &\approx& \epsilon. 
\end{IEEEeqnarray}

\paragraph*{Non-asymptotic approach}
While the asymptotic approach provides a simple closed-form expression for choosing the rate-selection function, asymptotic approximations are generally not accurate for small $l$. 
We can provide a simple, but less insightful, approximate bound for the meta-probability that does not exploit asymptotic normality as follows
\begin{IEEEeqnarray}{rCl}\label{eq:metaprob_ub_powerlaw}
\widetilde p_F &=& \mathbb{P}\left[{\widehat F_Z^{-1}(\varepsilon_n) > F_Z^{-1}(\epsilon)}\right]\\
&\leq& \min_{t}\mathopen{}\Bigg\{ \mathbb{P}\left[{Z_{(l)} > t}\right] +\mathbb{P}\left[{\sum_{i=1}^l (Z_{(l)} - Z_{(i)}) > \frac{l \left(\kappa \log \frac{\epsilon}{\alpha}-t  \right)}{\log\farg{\frac{n\varepsilon_n}{m}}}}\right]\Bigg\}\label{eq:evt_bound1}\IEEEeqnarraynumspace\\
&\approx& \min_{\tau\in[\epsilon, 1  ]}\Bigg\{2-I_{\tau}(l, n+1-l) - \frac{\gamma\left(l-1, \frac{l \log \frac{\epsilon}{\tau}}{\log\farg{\frac{n\varepsilon_n}{l}}} \right)}{(l-2)!}\Bigg\}.\label{eq:evt_bound2}
\end{IEEEeqnarray}
Here, \eqref{eq:evt_bound1} follows because, for any two random variables $A$ and $B$, $\mathbb{P}\left[{A + B > v}\right] \leq \min_t\mathopen{}\big\{\mathbb{P}\left[{A > t}\right] + \mathbb{P}\left[{B > v-t}\right] \}$. The step \eqref{eq:evt_bound2} follows from the substitution $\tau=\alpha \e{t/\kappa}$ and because, under the power law approximation, $\sum_{i=1}^q (Z_{(l)} - Z_{(i)})$ is Erlang distributed with parameters $1/\kappa$ and $l-1$ and $\mathbb{P}\left[{Z_{(l)} \leq t}\right] = \mathbb{P}\left[{U_{(l)} \leq F_Z(t)}\right]  \approx I_{\tau}(l,n+1-l)$ where $U_{(l)}$ is the $l$-th order statistic of a sequence of $n$ independent standard uniform random variables as in Section~\ref{sec:nonparametric}.

By choosing $\varepsilon_n$ such that right-hand side of \eqref{eq:evt_bound2} equals $\xi$, we find that 
\begin{IEEEeqnarray}{rCl}
	\widetilde p_F \lessapprox \xi. \label{eq:evt_bound3}
\end{IEEEeqnarray}
The approximation in \eqref{eq:evt_bound3} again originates from the power law approximation in \eqref{eq:power_approximation2}.

\section{Numerical evaluation}\label{sec:results}
\begin{figure*}[t]
\centering
\subfloat[$\epsilon=10^{-3},\xi=10^{-3}$]{\includegraphics[scale=0.73]{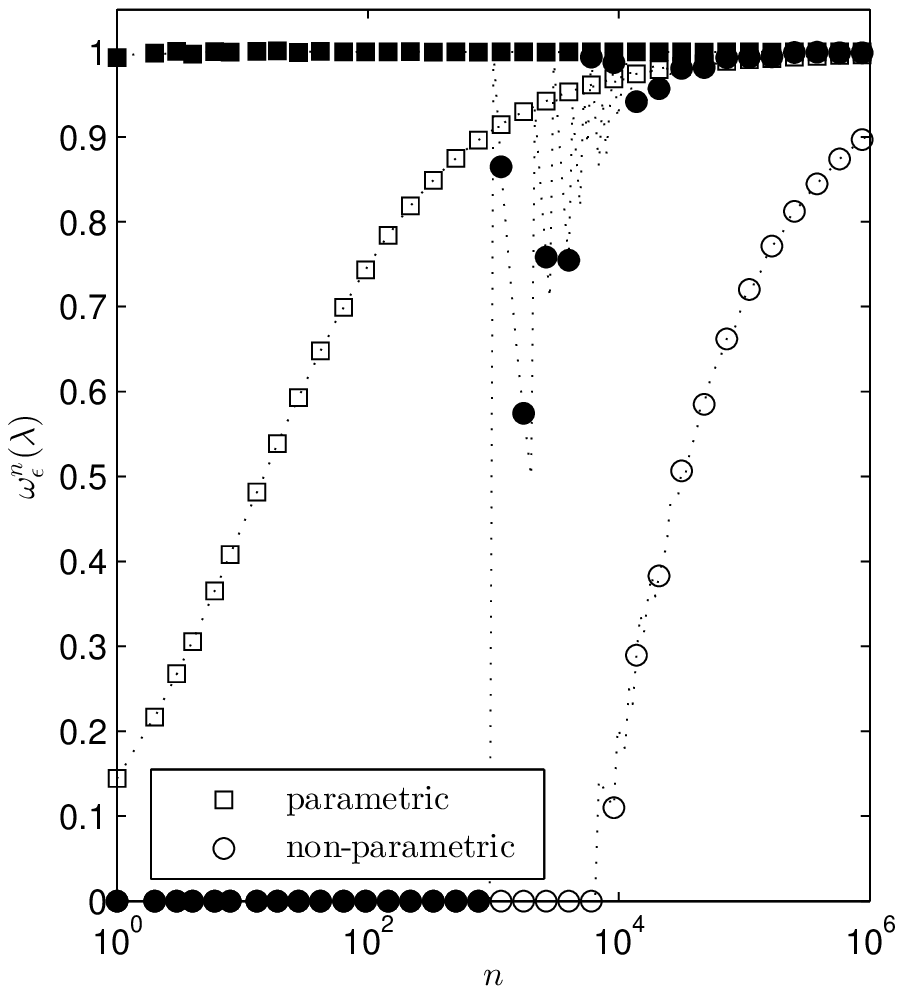}\label{results1a}}
\hfil
\subfloat[$\epsilon=10^{-4},\xi=10^{-3}$]{\includegraphics[scale=0.73]{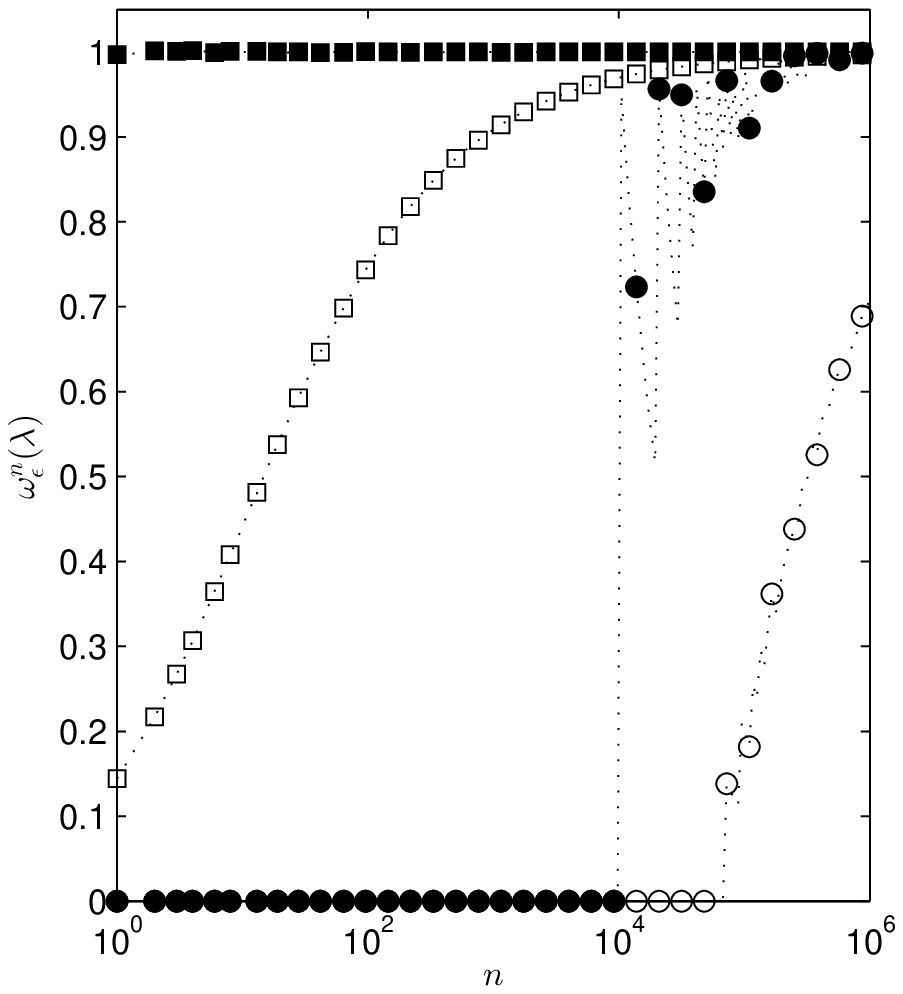}\label{results1b}}
\hfil
\subfloat[$\epsilon=10^{-5},\xi=10^{-3}$]{\includegraphics[scale=0.73]{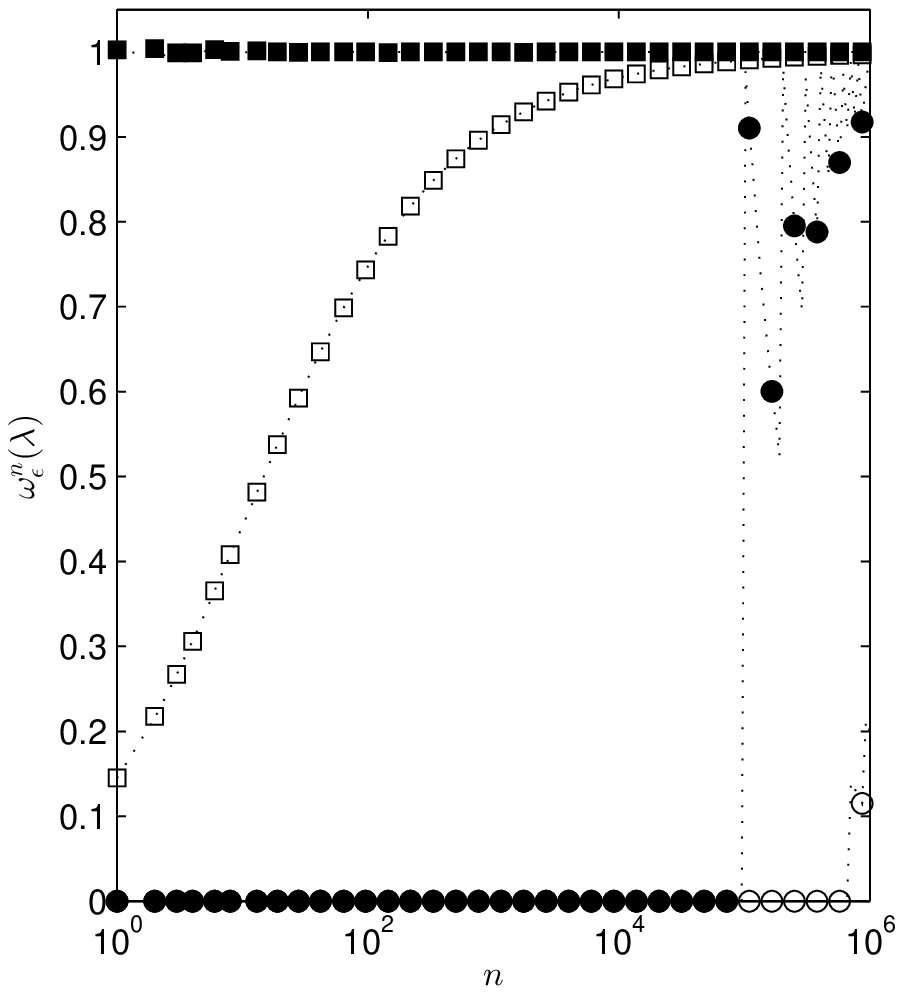}\label{results1c}}
\caption{Parametric vs non-parametric rate-selection under Rayleigh fading with $\lambda=1$ (full/empty markers correspond to mean outage probability/meta-probability constraints \eqref{eq:constI}/\eqref{eq:constII}), respectively. }
\label{SpecifiedParamAndNonparam_Rayl}
\end{figure*}

Before presenting the result, we  advise the reader to refer to the caption of the individual figures for details regarding the notation of the curves.

We begin by considering transmission over Rayleigh flat fading channel with average received power $\lambda=1$.
Fig.~\ref{SpecifiedParamAndNonparam_Rayl} shows the throughput ratio $\omega_{\epsilon}^n(\lambda)$ as defined in \eqref{eq:average_th_ratio_def} for parametric \eqref{eq:rate_selection_function_param} (square markers) and non-parametric (circle markers) rate-selection functions \eqref{eq:rate_selection_function_nonparam} for different training sample lengths $n$ and $\epsilon\in\left\{10^{-3},10^{-4},10^{-5}\right\}$.
We observe that the parametric and non-parametric rate-selection functions 
are consistent; 
the oscillations in the non-parametric case arise due to $(l)\in\mathbb{N}$.
We also observe that the parametric rate-selection functions converge significantly faster to the $\epsilon$-outage capacity.
As expected, the non-parametric rate-selection functions require $n$ to be of the order $1/\epsilon$ to produce non-zero throughput; the meta-probability constraint requires $n$ to be even larger, almost an order of magnitude larger than the mean outage probability constraint.

\begin{figure}[t]
\centering
\includegraphics[scale=0.73]{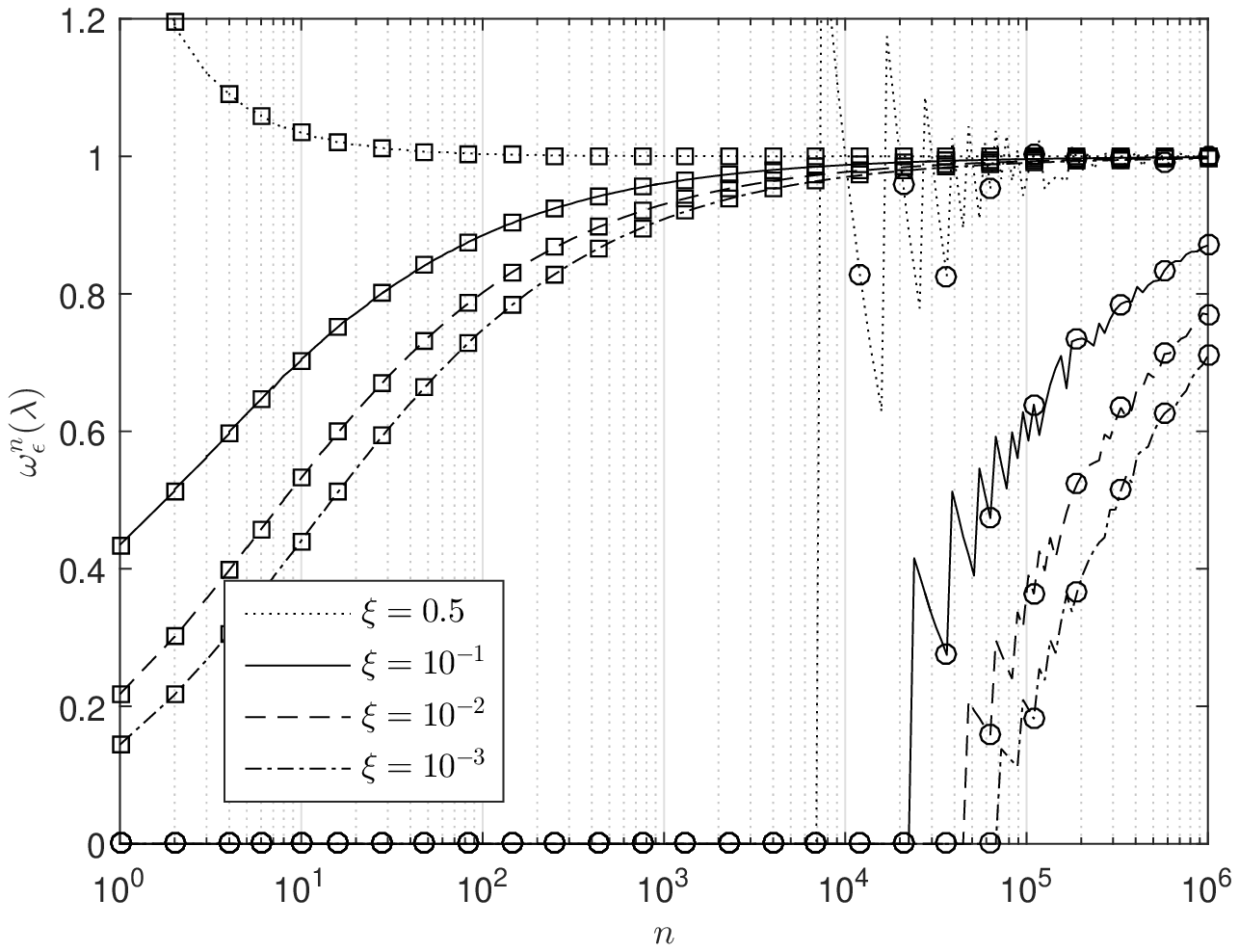}
\caption{Parametric rate-selection under Rayleigh fading and meta-probability constraint $\lambda = 1$, $\epsilon=10^{-4}$ (square/circle markers correspond to parametric/non-parametric rate-selection functions, respectively).}
\label{SpecifiedParamAndNonparam_Rayl_meta_diffxi}
\end{figure}

We observe that $\omega_\epsilon^n(\lambda)$ approaches $1$ faster when the rate-selection follows the mean outrage probability constraint \eqref{eq:constI} (filled markers) as opposed to meta-probability constraint \eqref{eq:constII} (empty markers); in fact, $\omega_\epsilon^n(\lambda)\approx 1$ and $R(x^n)\approx R_{\epsilon}(\lambda)$ even for $n<10$.
This is expected since the mean outage probability depends very weakly on $n$ (see \eqref{eq:mean_outage_weak_n}).
In contrast, the rate under meta-probability constraint converges slower which is also intuitively expected due to the strictness of the constraint. 
An interesting observation follows from Fig.~\ref{SpecifiedParamAndNonparam_Rayl}: the convergence rate under the parametric rate-selection function appears to be (almost) independent from $\epsilon$ for fixed $\xi$, implying that the rate-selection function obtained for given $\xi$ via \eqref{eq:varen_constII_rayleigh} is valid for any $\epsilon$.  
In Fig.~\ref{SpecifiedParamAndNonparam_Rayl_meta_diffxi} we depict the throughput ratio under meta-probability constraint for different values of $\xi$ (square/circle markers correspond to parametric/non-parametric rate selection).
Note that lower values for $\xi$ impose stricter requirements; this implies lower rate for fixed $n$ and slower convergence.
Interestingly, when $\xi\rightarrow 0.5$ we obtain similar behavior as in the case of mean outage probability.
In fact, when the distribution of the outage probability is symmetric, i.e., its mean coincides with the median, the meta-probability constraint for $\xi=0.5$ is equivalent to the mean outage probability constraint.
Hence, even though rigorously precise only in the case of symmetric outage probability distribution,  the meta-probability constraint \eqref{eq:constI} can, in general, be viewed as generalization of the mean outage probability constraint \eqref{eq:constII}. 

\begin{figure*}[t]
\centering
\subfloat[True distribution: Rician fading ($\lambda = 1$)]{\includegraphics[scale=0.33]{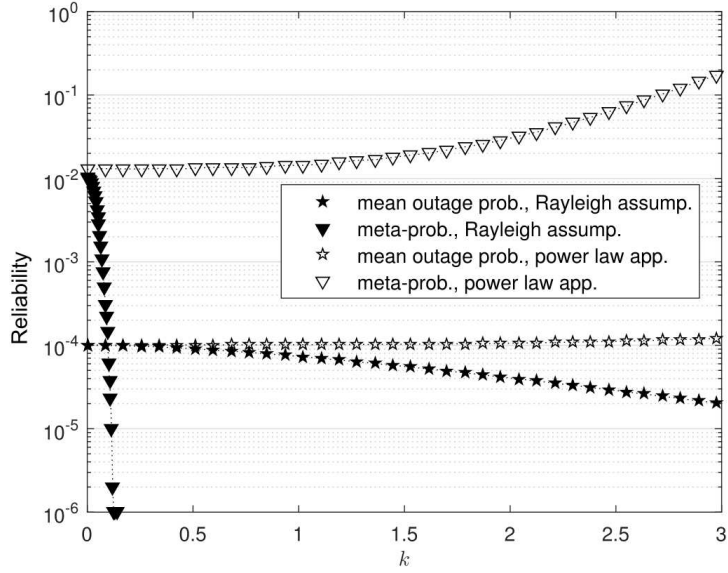}\label{results3a}}
\hfil
\subfloat[True distribution: Nakagami-$m$ fading ($\lambda = 1$)]{\includegraphics[scale=0.33]
{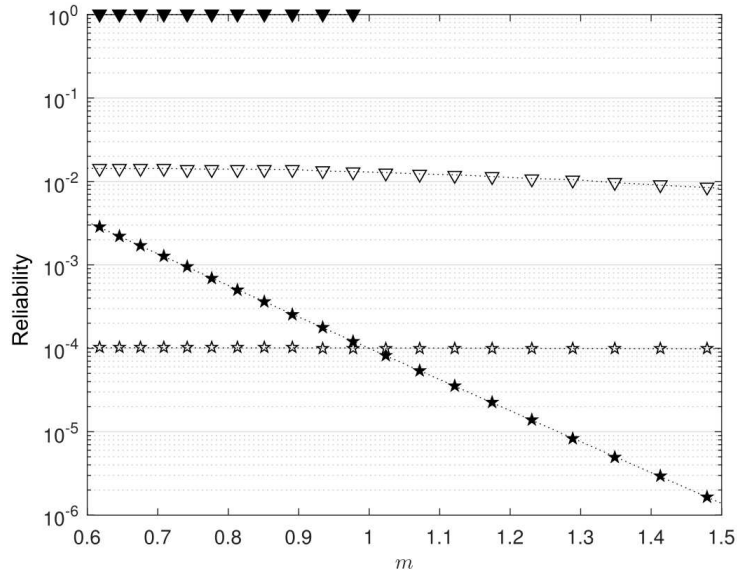}\label{results3b}}
\caption{Impact of channel mismatch onto mean outage probability and meta-probability with $\epsilon=10^{-4},\xi=10^{-2},n=10^6,\beta=0.01$ (full/empty markers correspond to Rayleigh fading assumption/power law tail approximation, respectively).}
\label{Missmatch}
\end{figure*}

Fig.~\ref{Missmatch} evaluates the impact of mismatched model on the reliability performance: the Tx adopts Rayleigh model (filled markers), but the true distribution is different.
We fix $\epsilon$ and $\xi$, we choose $n$ to be large enough to ensure convergence and we plot the mean outage probability and the meta-probability as a function of the true channel distribution as specified by the corresponding parameters.
In Fig.~\ref{results3a}, the channel follows Rice distribution \eqref{eq:cdf_X_rice} and we plot the reliability performance for a range of $k$-factors: note that, we plot both \eqref{eq:mean_outage_miss}/\eqref{eq:metaprob_miss} with the corresponding approximations \eqref{eq:mismatch_appI_1}/\eqref{eq:mismarch_appII_1} (dotted lines), confirming that the latter approximate the former well for small $\epsilon$ and $\xi$.
As already discussed in Section~\ref{sec:mismatched}, assuming Rayleigh when the actual fading is Rician, always gives a pessimistic rate-selection function, i.e., the throughput ratio is strictly less than $1$; in fact, as the specular component becomes stronger, the throughput becomes even lower.
We conclude that by assuming Rayleigh, i.e., $k=0$ when $k>0$, the reliability constraints \eqref{eq:constI} and \eqref{eq:constII} will never be violated.
However, this comes at the price of under-utilizing the degrees of freedom offered by the Rician channel; under meta-probability constraint, the underutilization is severe as seen in Fig.~\ref{results3a}, with the meta-probability quickly dropping and pushing the throughput towards $0$.
For Fig.~\ref{results3b} the true distribution is Nakagami-$m$ \eqref{eq:cdf_X_Nak} and it shows  the reliability performance for different values of $m$.
For $m>1$, the behavior is similar to the Rician case.
However, the numerical evaluations suggest that the under-utilization is more severe; already for $m=2$, which corresponds to two Rayleigh-type of clusters, both the mean outage probability and the meta-probability are several order of magnitudes below the thresholds, implying that the throughput is very low.
We observe opposite behavior for $m<1$ (which we also predicted in Section~\ref{sec:mismatched}).
Here, the rate-selection function gives optimistic rates, larger than the corresponding $\epsilon$-outage capacity; such rates violate the reliability constraints to produce throughput ratio larger than $1$.
Once again, the meta-probability shows severe under/over-utilization, quickly dropping/jumping towards $0/1$.

\begin{figure}[t]
\centering
\includegraphics[scale=0.73]{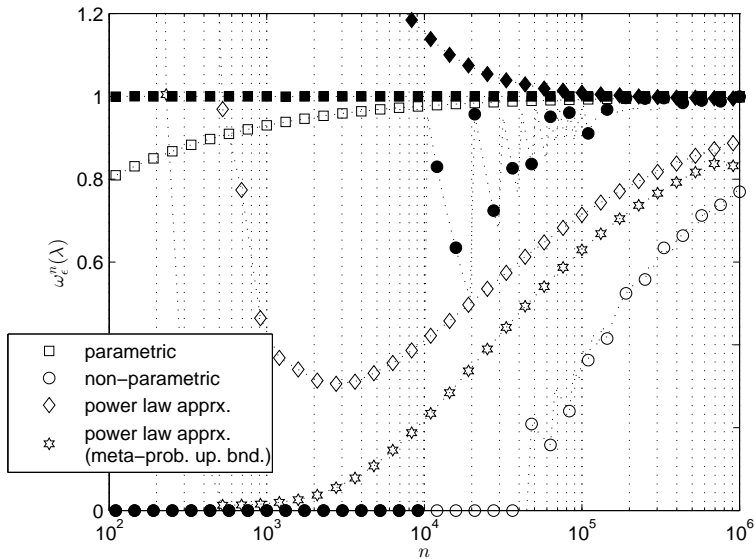}
\caption{Parametric, non-parametric and power law-based rate-selection under Rayleigh fading (full/empty markers correspond to mean outage probability/meta-probability, $\epsilon = 10^{-4}, \xi= 10^{-2}, \beta = 0.01, \lambda=1$).}
\label{ParamNonparamApp_Rayl}
\end{figure}

{The performance of the parametric and non-parametric rate-selection functions can be also linked to the bias-variance trade-off.
Namely, learning parametric models via MLE produces estimates with small variance which explains the relatively fast convergence of the rate, but they can be heavily biased in case of model mismatch. On the other hand, in the non-parametric approach the $\varepsilon_n$-quantile is estimated with a small bias but a large variance; hence, it works for any distribution $F$ but requires a large number of samples.}

The asymptotic rate-selection functions based on power law tail approximation address the trade-off among consistency, training sample length, and mismatch.
Fig.~\ref{ParamNonparamApp_Rayl} compares the performance of the power law tail approach against parametric/non-parametric rate-selection functions with Rayleigh true distribution; in such case $\kappa = 1$, $\alpha = 1/\lambda$.
Clearly, the throughput ratio converges faster than the non-parametric case, which is particularly evident for the meta-probability constraint.
Note that the approach requires $l\geq 2$, i.e., $n\geq 2/\beta$ samples.
Also, for small $l$ (less than $10$ in this case study), the estimators \eqref{eq:kappa_est} and \eqref{eq:alpha_est} over-estimate $\kappa$ and $\alpha$.
For the meta-probability constraint, this effect is (partially) alleviated by using $\varepsilon_n$ chosen according to the upper bound \eqref{eq:metaprob_ub_powerlaw} at the expense of slower convergence.
The rate-selection function \eqref{eq:rate_selection_powerlaw} also suffers from mismatch due to approximation error; therefore, the approach does not guarantee consistency.

To investigate the effect, Fig.~\ref{Missmatch} shows the reliability performance of the approach (empty markers), compared against the case of incorrect model assumption. 
In both Figs.~\ref{results3a} and ~\ref{results3b}, we see that even in Rayleigh case ($k=0$ or $m=1$, recall the special cases of the Rician or Nakagami-$m$ channel in Section~\ref{sec:prelim}) there is slight bias in the reliability performance which is more evident for the meta-probability constraint; hence, the rate is slightly inconsistent due to the approximation error.
When the true distribution is Ricean, we see that the power law tail approach violates both constraints for large $k$, producing optimistic rates; similarly, in the case of Nakagami-$m$ and $m<1$, the approach gives optimistic rates.
This is due to the fact that the power law tail approximation is a lower bound of the respective cdfs (see \cite{Eggers2017} for more details).
Oppositely, for large enough $m>1$ in the Nakagami-$m$ channel, the approach leads to pessimistic rates since the power law approximation becomes an upper bound \cite{Eggers2017}.

\section{Conclusions}\label{sec:conc}
The strict requirements of URLLC systems demand from us to reconsider the usual ways the physical layer is designed and its performance assessed.
Our study shows that the phenomena of channel uncertainty strongly impact the performance in UR-relevant regime of operation, in the sense that the targeted reliability can no longer be guaranteed with certainty.
Motivated by this, we introduced novel statistical framework in which the transmitter, given its best knowledge of the true channel distribution, determines the physical layer parameters, namely the transmission rate such that the reliability is guaranteed probabilistically.
We showed that the approach based on parametric channel models requires the least amount of channel training but is highly susceptible to modeling mismatch.
Non-parametric approaches do not suffer from such mismatch since they do not impose assumptions; however, the training sample sizes tend to be prohibitively large.
As a consensus between the parametric and non-parametric approaches, the design approach based on the power law tail approximations offers versatile performance that reduces both, the impact of modeling mismatch and the training sample sizes.

The main lesson from our study is the implication that attaining and guaranteeing the strict performance requirements in URLLC-relevant regimes of operation solely by adequate design of the physical layer is a challenging task.
The reader should note that apart from link outages, our study does not consider other effects such as noise, interference or equipment imperfections; combined, all of these effects would make it even more challenging to guarantee the reliability performance.
All of this suggest that the physical layer design in URLLC system should incorporate more advanced technologies that rely on diversity such as multiple antenna techniques which is part of our on-going research. 


\appendices
\section{Proof of \eqref{eq:meta_prob_asymp}}\label{app:powerlaw}
We are going to show that, under the power law approximation,
\begin{IEEEeqnarray}{rCl}
\sqrt{n}\left(\widehat F_Z^{-1}(\varepsilon_n) -\kappa \log \frac{\epsilon}{\alpha} + \sqrt{\frac{\kappa^2\overline V}{n}} Q^{-1}(\xi)\right)\label{eq:asymp_norm_prop}
\end{IEEEeqnarray}
is asymptotically normal with asymptotic mean and asymptotic variance given by $0$ and $\kappa^2 \overline V$, respectively. Let $W$ be a standard normal random variable. Then, we obtain \eqref{eq:meta_prob_asymp} through the following steps
\begin{IEEEeqnarray}{rCl}
    \IEEEeqnarraymulticol{3}{l}{\lim_{n\rightarrow \infty} \mathbb{P}\left[{\mathbb{P}\left[{R(X^n) \geq \log(1+Y) |X^n }\right]\geq \epsilon}\right]}\nonumber\\
    &=& \lim_{n\rightarrow \infty} \mathbb{P}\left[{\mathbb{P}\left[{\widehat F_Z^{-1}(\varepsilon_n)  > \log(Y) |X^n }\right]\geq \epsilon}\right]\\
    &=& \lim_{n\rightarrow \infty} \mathbb{P}\left[{\widehat F_Z^{-1}(\varepsilon_n) > F_Z^{-1}(\epsilon)}\right]\label{eq:evt_approx}\\
    &\approx&  \mathbb{P}\left[{ W > \sqrt{\frac{n}{ \overline V}}\left(\frac{  F_Z^{-1}(\epsilon)}{\kappa} - \log \frac{\epsilon}{\alpha}+\sqrt{\frac{ \overline V}{n}}Q^{-1}(\xi) \right)}\right]\IEEEeqnarraynumspace\label{eq:asymp_norm_used} \\
    &\approx&  \mathbb{P}\left[{ W > Q^{-1}(\xi)}\right]\label{eq:asymp_norm_used2} \\
    &=& \xi.
\end{IEEEeqnarray}
Here, \eqref{eq:evt_approx} follows from the log-transformation which implies that $\mathbb{P}\left[{\cdot \geq \log(Y)}\right] = F_Z(\cdot)$,  \eqref{eq:asymp_norm_used} follows from the asymptotic normality of \eqref{eq:asymp_norm_prop}, and \eqref{eq:asymp_norm_used2} follows because $F_Z^{-1}(\epsilon) \approx \kappa \log \frac{\epsilon}{\alpha}$.

To establish  asymptotic normality of \eqref{eq:asymp_norm_prop} under the power law approximation, we first substitute the expressions for $\widehat F_Z^{-1}(\varepsilon_n)$ and $\varepsilon_n$ (see \eqref{eq:hatFZ} and \eqref{eq:asymp_vareps}) and rewrite \eqref{eq:asymp_norm_prop} as follows
\begin{IEEEeqnarray}{rCl}
&&{\sqrt{n}\Bigg(Z_{(l)}- \kappa \log \frac{\beta}{\alpha}} \nonumber\\
&&\qquad {}+ \frac{1}{l}\log\farg{\frac{n\epsilon}{l}}  \sum_{i=1}^l (Z_{(l)}-Z_{(i)})-\kappa \log \frac{\epsilon}{\beta}\nonumber\\
&& \qquad{}-  \sqrt{\frac{\overline V}{n} } Q^{-1}(\xi)  \left(\frac{1}{l}\sum_{i=1}^l (Z_{(l)}-Z_{(i)})- \kappa \right)\Bigg).\IEEEeqnarraynumspace\label{eq:expanded_asymp_norm}
\end{IEEEeqnarray}
We now consider each line of \eqref{eq:expanded_asymp_norm} separately. 
First, it follows from \cite[Th.~21.7]{Vaart}  that
\begin{IEEEeqnarray}{rCl}
\sqrt{n}\bigg(Z_{(l)} - \kappa \log \frac{\beta}{\alpha}\bigg) &\stackrel{\text{d}}{\rightarrow}&  \mathcal{N}\farg{0, \widetilde V} 
\end{IEEEeqnarray}
as $n\rightarrow \infty$, where 
\begin{IEEEeqnarray}{rCl}
 \widetilde V &=& \frac{\beta(1-\beta)}{f_Z^2(F_Z^{-1}(\beta))} \approx  \frac{\kappa^2(1-\beta)}{\beta}.
\end{IEEEeqnarray}
Next, it follows from the standard central limit theorem \cite[Th.~2.17]{Vaart} that 
\begin{IEEEeqnarray}{rCl}
&&\sqrt{n}\Bigg( \frac{1}{l-1}\log\farg{\frac{\epsilon}{\beta}}  \sum_{i=1}^l (Z_{(l)}-Z_{(i)}) -\kappa \log \frac{\epsilon}{\beta} \Bigg)\label{eq:clt_used}
\end{IEEEeqnarray}
is asymptotically normal with mean $0$ and variance $\frac{\kappa^2}{\beta} \log^2 \frac{\epsilon}{\beta}$ (recall that the mean and variance of an Erlang distribution with parameters $1/\kappa$ and $l-1$ is $\kappa(l-1)$ and $\kappa^2(l-1)$, respectively. An application of Slutsky's theorem \cite[Th.~2.8]{Vaart} shows that the second line in \eqref{eq:expanded_asymp_norm} (multiplied by $\sqrt{n}$) has the same asymptotic distribution as \eqref{eq:clt_used}. We finally note that
\begin{IEEEeqnarray}{rCl}
\frac{1}{l}\sum_{i=1}^l (Z_{(l)}-Z_{(i)})
\end{IEEEeqnarray}
converges to $\kappa$ almost surely as $n\rightarrow \infty$.

These three properties imply that \eqref{eq:asymp_norm_prop} is asymptotically normal distributed with asymptotic mean and asymptotic variance given by $0$ and $\frac{\kappa^2}{\beta}(1-\beta + \log^2 \frac{\epsilon}{\beta}) = \kappa^2  \overline V$, respectively, as desired.\footnote{Note that if $A_n \stackrel{\text{d}}\rightarrow \mathcal{N}(0,\sigma_1^2)$ and $B_n \stackrel{\text{d}}\rightarrow \mathcal{N}(0,\sigma_2^2)$ as $n\rightarrow \infty$, then it follows from L\'evy's continuity theorem \cite[Th.~2.13]{Vaart} that $A_n + B_n\rightarrow \mathcal{N}(0,\sigma_1^2+\sigma_2^2)$.}





\ifCLASSOPTIONcaptionsoff
  \newpage
\fi

\bibliographystyle{IEEEtranTCOM}


\end{document}